\documentclass[aps,preprint,pre,showpacs,amssymb,superscriptaddress]{revtex4-1}

\usepackage{graphicx}
\usepackage{bm}
\usepackage{epsfig,amsmath, amsfonts, amssymb, xcolor,hyperref,wasysym}
\usepackage{ stmaryrd}
\usepackage{soul}
\usepackage{textcomp}

\begin{document}

\title{ Critical Casimir forces in the presence of random surface fields}

\author{A.~Macio\l ek}
\email{maciolek@is.mpg.de}
\thanks{ corresponding author}
\affiliation{Max-Planck-Institut f{\"u}r Intelligente Systeme, Heisenbergstr.~3, D-70569 Stuttgart, Germany}
\affiliation{IV. Institut f\"ur Theoretische  Physik,
  Universit\"at Stuttgart, Pfaffenwaldring 57, D-70569 Stuttgart, Germany }
\affiliation{Institute of Physical Chemistry,
             Polish Academy of Sciences, Kasprzaka 44/52,
            PL-01-224 Warsaw, Poland}
\author{O. Vasilyev}
\affiliation{Max-Planck-Institut f{\"u}r Intelligente Systeme, Heisenbergstr.~3, D-70569 Stuttgart, Germany}
\affiliation{IV. Institut f\"ur Theoretische  Physik,
  Universit\"at Stuttgart, Pfaffenwaldring 57, D-70569 Stuttgart, Germany }
\author{V. Dotsenko}
\affiliation{LPTMC, Universit´e Paris VI - 75252 Paris, France}
\affiliation{L.D.\ Landau Institute for Theoretical Physics,
           119334 Moscow, Russia}
\author{S. Dietrich}
\affiliation{Max-Planck-Institut f{\"u}r Intelligente Systeme, Heisenbergstr.~3, D-70569 Stuttgart, Germany}
\affiliation{IV. Institut f\"ur Theoretische  Physik,
  Universit\"at Stuttgart, Pfaffenwaldring 57, D-70569 Stuttgart, Germany }

\date{\today}


\begin{abstract}  We study  critical Casimir forces (CCF) $f_{\mathrm C}$  for films  of thickness $L$  which in the three-dimensional bulk belong to the Ising universality class
and which are exposed to random surface fields (RSF) on both surfaces. We consider the case that, in the absence of RSF, the  surfaces of the film  belong  to the surface universality class of the so-called ordinary transition.
 We  carry out a  finite-size scaling analysis and show that  for weak disorder  CCF  still  exhibit scaling, acquiring  a random field  scaling variable $w$ which  is zero for pure systems.
We confirm these  analytic predictions  by  MC simulations.   Moreover, our  MC data show that  $f_{\mathrm C}$ varies as $f_{\mathrm C}(w\to 0)-f_{\mathrm C}(w=0)\sim w^2$. Asymptotically, for large $L$, $w$ scales as 
$w \sim L^{-0.26} \to 0$ indicating that this type of disorder is an irrelevant perturbation of the  ordinary surface universality class. However, for thin films such that $w \simeq 1$, 
we find that the presence  of RSF with vanishing mean value  increases significantly the strength   of CCF,  as compared to  systems without them, and shifts the extremum  of the scaling function of $f_{\mathrm C}$  towards lower temperatures. But $f_{\mathrm C}$  remains attractive. 
\end{abstract}

\maketitle

\section{Introduction}
\label{sec:intr}

Critical Casimir forces (CCF)  arise between  surfaces confining  a fluid which is 
brought thermodynamically close to its bulk critical point \cite{FdG}.
 There they are described by universal scaling functions which are  determined by the universality class of 
the bulk liquid and  the surface universality classes of the  confining surfaces \cite{binder}.
Interfaces confining $^4$He near its superfluid transition belong to the surface universality 
class of the so-called ordinary transition  corresponding to  Dirichlet boundary conditions (BC) for the superfluid order parameter \cite{chan}.
Surfaces confining classical binary liquid mixtures near their demixing transition belong to the universality class of the
so-called normal  transition
 \cite{pershan,rafai,nature,nature_long,nellen}, which is characterized by
 a strong effective surface field acting on the deviation of the concentration from its critical value serving 
 as the order parameter. 
 The surface field describes the preference of the container wall
for one of the two species forming the binary liquid mixture. For $^3$He/$^4$He mixtures near their tricritical
point both types of BC can occur \cite{chan1}.
These experimental findings agree with corresponding theoretical analysis  \cite{dietrich_krech,krech,upton,md} and Monte Carlo
simulations \cite{VGMD,hucht,Hasenbusch}
of suitable model systems representing the aforementioned universality classes and the crossover between them \cite{MMD,abraham_maciolek,vas-11,diehl}. 
 Across the various universality classes the magnitude, the shape, and the sign  of the universal scaling functions of the CCF vary strongly. 
For example, for films with  ordinary-ordinary $(o,o)$ or normal-normal $(+,+)$ 
 BC at the  two surfaces  CCF  are attractive  CCF   whereas for   opposing $(+,-)$ BC they are  repulsive.

The sign of the surface fields depends on the chemical composition of the wall surfaces. 
They can be designed by suitable surface treatments which, e.g., render hydrophilic or hydrophobic surfaces  \cite{nature,nature_long}.
 In the context of CCF spatially varying surface compositions have been studied experimentally for a smooth lateral gradient \cite{NHB}
 and for well defined alternating stripes \cite{TZGVHBD}. Without dedicated preparation efforts, the surfaces typically carry random chemical heterogeneities 
due to adsorbed impurities which act as local surface fields. If kinetically frozen they form quenched disorder. 
Quenched random-charge disorder on surfaces of dielectric parallel walls at a distance $L$ leads to long-ranged forces $\sim L^{-2}$ 
even if they are net neutral \cite{NDSHP,BAD}, which dominates the pure van der Waals term  $\sim L^{-3}$.

Here we study CCF emerging under the influence of randomly quenched surface fields. 
Specifically, we consider the Ising bulk universality  class and a situation in which the mean value of the surface fields
vanishes. As a rough guideline this addresses systems in which droplets
 of the demixed binary liquid mixture form a contact angle of $90^0$ with the chemically disordered substrate 
(see the intermediate substrate compositions in Ref.~\cite{NHB}). We analyze slabs of thicknesses $L$. 
In the corresponding limit $L\to \infty$, 
leading to two semi-infinite systems, the influence 
of random surface fields has been studied in the context  of wetting (for reviews see Ref.~\cite{dietrich}) and surface
critical phenomena  \cite{mon_nig,diehl_nusser1,cardy}
  (for a review see Ref.~\cite{Pleimling}). In particular,  the  Harris criterion concerning the relevance of disorder for bulk critical 
phenomena has been generalized to  surface critical behavior
\cite{diehl_nusser1}. Within the framework and limitations of a weak-disorder expansion,
 quenched random  surface fields with vanishing mean value are   expected to be irrelevant if the pure
 system belongs to the ordinary surface universality class  \cite{diehl_nusser1}.
For the  three-dimensional ($d=3$) Ising model in Ref.~\cite{mon_nig} this  was  pointed out and confirmed by Monte Carlo simulations. 

Parallel to the present study, in Ref.~\cite{francesco}  the case of random surface fields acting on
only  one of the two  confining surfaces,  with  the other
surface belonging to the  universality class of the  normal transition, 
has been analyzed  for the  'improved'
 Blume-Capel model~\cite{Blume,capel,Hasenbusch}.
 The scaling functions of  CCF  in that system   have been  obtained by using Monte Carlo simulations and  finite-size scaling.
We note   that for  complex fluids   disorder effects on  
Casimir-like interactions can be  dominant. This has been shown recently for  nematic liquid-crystalline films
bounded by two planar surfaces, one of which  exhibiting  a random distribution of 
the preferred anchoring axis   in the so-called easy direction \cite{podgornik}.
In this  case of quenched disorder, the effects of disorder onto the 
fluctuation-induced interaction between the surfaces are dominant at intermediate film thicknesses.

Our presentation is organized as follows. In Sec.~\ref{sec:rs} we present a scaling analysis from which we derive a
 random field finite-size
scaling variable. Section ~\ref{sec:mc} is devoted to  MC simulations. In Sec.~\ref{subsec:mod}  
we define our system and  provide the details of our numerical  method of determining the CCF 
and its scaling functions 
from the MC simulation data. 
Section \ref{subsec:numres} contains our results. We provide a summary and conclusions in Sec.~\ref{sec:concl}.

\section{Random field scaling}
\label{sec:rs}

First, we  consider  pure systems.  Within  mean field  theory, near the ordinary transition of semi-infinite 
systems the order parameter profile exhibits  an extrapolation 
length $1/c$; $c=\infty$ is the fixed point of  the ordinary transition (o)  corresponding to Dirichlet BC~\cite{binder}. 
Close to  this ordinary transition there is a single linear scaling field $g_{1}=H_1/{\tilde c}^{y_c}$
associated  with the dimensionless surface 
field of strength $H_1$ and the surface enhancement parameter $\tilde c=ca$, where $a$ is a characteristic 
microscopic length scale of the system
  \cite{binder} such as the amplitude $\xi_0^{\pm}$ of the bulk correlation length
  $\xi_b(t=\frac{T-T_c}{T_c}\to 0^{\pm}) \;\simeq \; \xi_0^{\pm}|t|^{-\nu}$ 
  (which stands for asymptotic equality).
In the following  all lengths, such as $L$ and $1/c$, are expressed in units of $a$.
The above scaling exponent is $y_c=\left(\Delta_1^{sp}-\Delta_1^{ord}\right)/\Phi$, 
 where  $\Delta^{ord}_{1}$ and $\Delta^{sp}_{1}$ are the surface counterparts at the ordinary and special transition, respectively, of the bulk gap exponent $\Delta$,
 and $\Phi$ is a  crossover exponent \cite{binder}.  Within mean field theory one has $y_c=1$ 
 whereas $y_c(d=3)\;\approx \; 0.87 $ ~\cite{binder,GZ,vas-11}.
Close to the critical point, the {\it sing}ular part $f_{sing}$ of the free energy per  $k_BT$ and   divided by $AL$
 of a film of thickness $L$  and the surface area $A$
depends on three (dimensionless)
scaling fields: $t$, the bulk ordering field $h_b$, and  $g_1$; it depends on $L$ but not on $A$.  For  $L \gg a$, it
is a generalized homogeneous function so that
  $f_{sing}(t,h_b,g_1;L^{-1};a)\; \simeq  \; b^{-d}f_{sing}(b^{y_t}t,b^{y_b}h_b,b^{y_1}g_1;bL^{-1};a$) for any dimensionless
   rescaling factor $b>0$ and  bulk spatial dimension $d\ge 2$.
The scaling exponents  $y_t$, $y_b$, and $y_1$  are  related to the aforementioned critical exponents:
$y_t=1/\nu$, $y_b=\Delta/\nu$, and $y_1=\Delta^{ord}_1/\nu$. 
 (Note that $L^{-1}$ can be treated as a scaling field with scaling exponent  equal to 1.)
Setting $b=L/a$ one obtains (omitting the nonuniversal amplitudes of the scaling fields)
$f_{sing}(t,h_b,g_1;L^{-1};a) \; \simeq  \; (L/a)^{-d}f_{sing}((L/a)^{1/\nu}t,(L/a)^{\Delta/\nu}h_b,(L/a)^{\Delta^{ord}_1/\nu}g_1;a;a)$.

We now consider a Gaussian distribution of surface fields with the ensemble averages $
\overline{H_{1}(\mathbf{r})}=0
$
and
$
\overline{H_{1}(\mathbf{r}) H_{1}(\mathbf{r}')} = h^{2}\delta(\mathbf{r} - \mathbf{r}');
$
$\mathbf{r}$ and $\mathbf{r'}$ denote dimensionless lateral positions. In this case, the above finite-size scaling relation for the free energy density is modified.
A heuristic renormalization group argument \cite{mon_nig,and_ber} predicts that
the scaling exponent  of  a random surface field is
$y_{1}-(d-1)/2$.
This argument is based on the assumption that in  a surface block of side length $b$ and area $b^{d-1}$
the effect, on the pure system, of small quenched local fluctuations 
of the surface field $H_1(\mathbf{r})$ of average magnitude $h$ and zero mean 
is the same as that of the average strength $\left(\overline{(H^{cg}_{1}(\mathbf{r}))^2}\right)^{1/2}$ of the
 {\it c}oarse {\it g}rained  random field $H_1^{cg}=\displaystyle\sum_{i=1}^{{\cal N}_b} H_1(\mathbf{r}_i)$ 
uniformly distributed over the ${\cal N}_b$ sites $\mathbf{r}_i$  of that   block.
Since  $H_1^{cg}$ is the sum of ${\cal N}_b$ uncorrelated random (surface) fields one has 
$\left(\overline{(H^{cg}_{1}(\mathbf{r}))^2}\right)^{1/2}\sim h \sqrt{{\cal N}_b}$. With ${\cal N}_b \sim b^{d-1}$ one obtains 
$\left(\overline{(H^{cg}_{1}(\mathbf{r}))^2}\right)^{1/2}/{\cal N}_b \sim h/\sqrt{{\cal N}_b} \sim hb^{-(d-1)/2}$.

A real space renormalization-group transformation replaces such a  block  by a single site of the renormalized
system  with the associated quenched fluctuation of strength
$ b^{y_{1}}b^{-(d-1)/2}h$. Thus in a system with random surface  fields the appropriate scaling variable, which  
replaces  $(L/a)^{\Delta^{ord}_1/\nu}g_1$ 
for the pure system,   is
\begin{equation}
 \label{eq:w}
 w \equiv  (L/a)^{\Delta^{ord}_1/\nu -(d-1)/2}h/c^{y_c} =  (L/a)^{y_1-(d-1)/2}{\tilde g_1},
\end{equation}
where ${\tilde g_1} = h/c^{y_c}$; in the following we choose $a = \xi_0^+$.
As in our previous study \cite{vas-11}, for the  three-dimensional $(d=3)$ Ising model we take  
 $\Delta^{ord}_{1} \; \approx \;  0.46(2)$  \cite{GZ},  $\Delta^{sp}_{1} \;  \approx \;  1.05$~\cite{binder}, 
$\Phi \;  \approx  \; 0.68$~\cite{binder}, 
 and $\nu \approx 0.63$ \cite{PV,Hasenbusch},
and obtain 
$y_{1}-(d-1)/2 \;  \approx \; -0.26(6)$.
(More accurate estimates for the surface critical exponents at the  special and  ordinary 
transitions were  obtained recently 
from MC simulations \cite{Hasenbusch84}. They yield  $y_c\approx 1.282(5)$
and $y_1\approx 0.7249(6)$ so that $y_{1}-(d-1)/2  \approx -0.2750(4)$.
We have checked that using these latter estimates does not change the  conclusion of our study and  
yields  very similar results.)
Within  
  mean field theory, i.e., for $d=4$, one has $\Delta_1^{ord} = \nu=1/2$ \cite{binder} so that $y_{1}-(d-1)/2 = -1/2$.
In the marginal case  $ d=2 $ one has   $ y_{1}-(d-1)/2 =0$, due to $\nu=1$ and $\Delta^{ord}_1=1/2$ \cite{binder}.
Accordingly, for the  $d=3$ Ising model one has $w=(h/c^{0.87})(L/\xi_0^+)^{-0.26}$
whereas within mean field theory $w= (h/c)(L/\xi_0^+)^{-1/2}$.

At vanishing bulk ordering field $h_b =0$, the singular part of the  excess free energy 
$f^{ex}_{sing}(L^{-1}) = f_{sing}(L^{-1})-f_{sing}(0)$ (per $k_BT$ and divided by $AL$) satisfies 
(see Eq.~(3.18) in Ref.~\cite{Wegner} and Eqs.~(1.7) and (1.8) in Ref.~\cite{privman})
\begin{eqnarray}
 \label{eq:fex}
 f^{ex}_{sing}(t,h_b=0,{\tilde g_1};L^{-1};\xi_0^+) & \simeq  & (L/\xi_0^+)^{-d}f^{ex}_{sing}((L/\xi_0^+)^{1/\nu}t,(L/\xi_0^+)^{y_1-(d-1)/2}{\tilde g_1};\xi_0^+,\xi_0^+) \\ \nonumber
 & = & L^{-d}\Theta(L/\xi_0^+)^{1/\nu}t,(L/\xi_0^+)^{y_1-(d-1)/2}{\tilde g_1}).
\end{eqnarray}
 Accordingly,  the critical Casimir force per area $A$ and in units of $k_BT$  defined as 
 \begin{equation}
 \label{eq:def_Cas}
  f_{\mathrm C} \equiv - \frac{\partial(Lf^{ex}_{sing})}{\partial L},
 \end{equation}
satisfies
\begin{equation}
 \label{eq:scal_force_0}
f_{\mathrm C}(T,L,h) \simeq L^{-d}\vartheta(x = L/\xi_0^+)^{1/\nu}t,w = (L/\xi_0^+)^{y_1-(d-1)/2}{\tilde g_1}).
\end{equation} 
As follows from Eq.~(\ref{eq:def_Cas}), the scaling function  $\vartheta$ is related to $ f^{ex}_{sing}$ and its derivatives.
Because the scaling exponent  of  the random surface field is negative, the scaling field  ${\tilde g_1}$ is irrelevant 
in the  sense of
renormalization-group theory.
Under the assumption that the scaling function $\vartheta$ can be  expanded in powers of the irrelevant field ${\tilde g_1}$, 
one obtains the critical Casimir force 
\begin{eqnarray}
  \label{eq:f_C_exp}
f_{\mathrm C}(T,L,h) &\simeq & L^{-d}\vartheta + {\tilde g_1}(\xi_0^+)^{y_1-(d-1)/2} L^{-d-(y_1-(d-1)/2)}\vartheta_1 \\ \nonumber
&+& {\tilde g_1}^2(\xi_0^+)^{2(y_1-(d-1)/2)} L^{-d-2(y_1-(d-1)/2)}\vartheta_2 + \ldots,
\end{eqnarray}
where $\vartheta_1$ is the first derivative of $\vartheta$ with respect to $w$ and $\vartheta_2$ is the second derivative.
In Eq.~(\ref{eq:f_C_exp}) the function $\vartheta$ and its
derivatives are evaluated at $w=0$.
For sufficiently  thick films the effect of  disorder
 is expected to be negligible, i.e., the second  variable in Eq.~(\ref{eq:scal_force_0}) can be neglected in the limit $L \to \infty$.
However, in $d=3$ the exponent $y_1-(d-1)/2$ is small and 
 for thin films the corrections to scaling due to ${\tilde g_1}$ (i.e., the second and perhaps also the third term in
 the expansion in Eq.~(\ref{eq:f_C_exp}))
 can be large and hence important for experimental realizations. For large ${\tilde g_1}$ and thin films, it may happen that in 
 Eq.~(\ref{eq:f_C_exp}) even   more terms 
 have to be included in order to capture the behavior of $f_{\mathrm C}(T,L,h)$.

In the pure case, there is a length 
$\ell_{ord} = gg_1^{-1/y_1}$ associated with the scaling field $g_1$ \cite{diehl:97}, where $g$ is a nonuniversal amplitude, which 
 can be small
or large depending on the relative strength  of $H_1$ and $\tilde c$. Upon approaching the ordinary transition, i.e., in the 
limit $c \to +\infty$ at fixed $H_1$  one has $\ell_{ord}\to \infty$. On the other hand, in the limit $H_1 \to \infty$ 
at fixed $c > 0$,  one has
$\ell_{ord} \to 0$ at 
the  normal transition $(+)$. In $d=3$ one has $\ell_{ord} = g_1^{-1.3793(9)}$.
Various studies  of  Ising systems in the film geometry  \cite{RC,CR,MCD,MEW,MDB,MMD,abraham_maciolek,vas-11} showed that close to the ordinary transition  
the critical properties of the film of thickness $L$ are particularly 
sensitive to the  strength of the surface fields, i.e., 
whether the length scale 
$\ell_{ord}$ becomes comparable to or even larger than  $L$,   where criticality means 
$L, \ell_{ord} \ll \xi_b$.
In particular, in 
films with identical surface fields, 
the absolute  value of the critical Casimir force 
at the bulk critical temperature (characterized by the critical Casimir amplitude)  as a function of the
surface field $H_1$  exhibits  a minimum at $L \simeq \ell_{ord}$  \cite{MCD,vas-11}.
For  equal surfaces, also the effect of  the variation of  the amplitude 
of $H_1$  on the temperature dependence of the critical Casimir force,
 i.e., the crossover behavior between the ordinary and  the normal surface
 universality classes,
 has been  studied \cite{MDB,MMD,vas-11}.
For  $L/\ell_{ord}  \approx  1$  these results  show  strong deviations of the force scaling
function  from its universal fixed-point  behavior 
such as the  occurrence
of two minima, one above and one below   $T_c$, but  {\it no}
 change in sign  as the
 temperature is varied.

In the case of disorder,  due to the scaling exponent $y_1-(d-1)/2$ of a random surface field 
one can identify a  length scale $\ell$ associated with the  latter as 
 $\ell = \kappa\left(h/c^{y_c}\right)^{-1/(y_1 -(d-1)/2)}$  where  
 $\kappa$ is a microscopic length.
 In $d=3$  one has $\ell = \kappa \left(h/c^{0.87}\right)^{3.85}$. Conversely to  $\ell_{ord}$ in
 the pure case, at the ordinary transition, i.e.,
 in the limit of $c\to +\infty$  at fixed $h$, $\ell$ vanishes.  One can also consider the limit of large $h$  at fixed $c>0$
 which, however, does not correspond to
 the normal transition. In the case of a random surface field, $h$ is a standard deviation of the Gaussian probability distribution 
 of the surface fields.
Upon increasing  $h$ the probability distribution  broadens  so that also
 strong  surface fields  occur.
 Since  the presence of a surface field of  strength 
 $|H_1| \gtrsim 1$ eliminates  the fluctuations of  the boundary  Ising spins,
we expect that for $h \gtrsim 1 $ the actual value of $h$  does no longer  matter 
and that  accordingly the variation of the free energy with
$h$  levels off. The  typical microscopic configuration of  a surface layer exposed
to the realization of  random surface fields with large standard deviation $h$ will be that
of  non-fluctuating spins distributed  spatially almost  at random. 
(Surface fields with $|H_1| \lesssim 1$ will give rise to some
correlation effects.)  For  the finite size of a surface  as used in numerical simulations, 
 fluctuations about the zero mean value of the random surface field are expected to result in a
 nonvanishing, albeit small, surface magnetization.
For thin films and for  suitably chosen $h$ and $c$ such that  $\kappa h/c^{0.87} > 1$,  $\ell$    can be  comparable to $L$ or even larger. 
It is an  interesting issue whether, in analogy to the pure case, 
near bulk criticality
the presence of the  length scale $\ell$, which competes with $L$,
has  important consequences for the critical Casimir force.  In other words, using the following
representation of Eq.~(\ref{eq:scal_force_0}), 
\begin{equation}
 \label{eq:scal_force_1}
f_{\mathrm C}(T,L,h) \simeq L^{-d}{\hat \vartheta}(L/\xi_b,L/\ell),
\end{equation}
where ${\hat \vartheta} = \vartheta\left((L/\xi_b)^{1/\nu},(L/\ell)^{1/(y_1-(d-1)/2)}\right)$, we pose  the question  whether for $L \lesssim \ell$ one can observe significant deviations of the force scaling 
function ${\hat \vartheta}$ from its universal  ordinary-ordinary $(o,o)$ fixed-point ($c=\infty, h=0$)  behavior
${\hat \vartheta}(L/\xi_b,L/\ell = \infty)$. 
 We address this question  in the following section by  using MC simulations.

\section{Monte Carlo simulations}
\label{sec:mc}

\subsection{The model and the method}
\label{subsec:mod}

We have performed MC simulations of an  Ising model on a cubic lattice of   size
$L_{x} \times L_{y} \times L_{z}  $ with  $L_{x}=L_{y}=6 L_{z}$.  Here and in the following all lengths are measured in units of the lattice
constant $a_0$.
The spins  $s_{i_x,i_y,i_z}=\pm 1$ are located at every  lattice site  with the coordinates $i=(i_x,i_y,i_z)$,
$1 \le i_x \le L_{x}$, $1 \le i_y \le L_{y}$, $1 \le i_z \le L_{z}$.
The Hamiltonian of this lattice  model is given by
\begin{equation}
\label{eq:Ham}
 {\mathcal H}=-J\left(
\sum \limits_{\left<nn \right>}s_{i}s_{i'}
+\sum_{i_x,i_y } H_{1}(i_x,i_y) s_{i_x,i_y,1}
+\sum_{i_x,i_y } H_{2}(i_x,i_y) s_{i_x,i_y,L_{z}}
 \right),
\end{equation}
where $J>0$ is the spin coupling  constant,
the sum $\left<nn \right>$ is taken over nearest neighbors,
and $H_{1}(i_x,i_y)$ and $H_{2}(i_x,i_y)$ are dimensionless  random   fields acting on 
the top and the bottom surface, respectively,  of the system. The surface fields are independent random variables with 
a Gaussian distribution, with vanishing  mean values
$\left< H_{1}(i_x,i_y)\right>=\left< H_{2}(i_x,i_y)\right>=0$,
and with  half-widths $
h^{2}=
\left< \left(H_{1}(i_x,i_y)\right)^{2}\right>
=\left< \left(H_{2}(i_x,i_y)\right)^{2}\right>.$
The computations have been performed  for systems with thicknesses $L_{z}=10,15$, and $20$. We have used the so-called
 coupling parameter  method
 in order to determine  the CCF $f_{\mathrm C}(\beta,L,h)$, where $\beta=J/(k_BT)$ is the reduced inverse temperature  and 
 $L=L_{z}-\frac{1}{2}=9.5,14.5$, and $19.5$ is the slab thickness (in units of the lattice spacing)
 corresponding to  the force $f_{\mathrm C}(\beta,L,h)$. This method has been  employed in  previous MC simulations determining
  the CCF
for pure films \cite{VGMD,vas-11}.
We  have used the following numerical properties of this Ising model:
$\beta_{c}=0.2216544(3)$ \cite{TB}, $\nu=0.6301(2)$ \cite{PV}, and $\xi_{0}^{+}=0.501(2)$ \cite{RZW} in units of the lattice spacing
$a_0$.

Since according to Eq.~(\ref{eq:Ham})  the coupling constant within  the surface layers 
and between the surface layers 
and their  neighboring layers is the same as in the bulk, the  corresponding surface enhancement is, within  mean-field theory and in units
of the lattice spacing,
$c=1$ \cite{binder}. Beyond mean-field theory, the relation between  $c$ and  the coupling constants is not known.
In order to proceed, in the following we set  $c^{0.87}=1$ and  use 
 the scaling variable $\hat w=h/L^{0.26}$.  Accordingly,  $\ell = h^{3.85}$ so that for the  thicknesses $L = 9.5, 14.5,$ and 19.5
 used here 
 the condition $\ell \simeq L$ is satisfied for  $h = 1.80, 2.00,$ and 2.16, respectively.
The value $\hat w=0$ corresponds to films with free $(o,o)$ BC.

For every value of the scaling variable  $\hat w$ we have performed  an ensemble  average  over $N_{r}=64$ independent
realizations $j=1,2,3,\dots,N_{r}$ of random  surface fields.
For every realization  of  systems  with  $L_{z}=10,15,20$ lattice layers
the thermal average  is performed over $10^{5}$, $5\times 10^{4}$, and $2.5\times 10^{4}$ 
 hybrid MC steps, respectively, split into  10
series in order to  determine the statistical error. 
We denote $f_{\mathrm C}^{j}(\beta,L,h)$ as the critical  Casimir force  per $k_BT_c$ and per surface area $S_2=L_x\times L_y$, obtained
from the  $j$-th realization at  the inverse temperature 
$\beta = 1/(k_BT)$, for the system thickness $L$ and for  the random surface field scaling variable 
$\hat w=h/L^{0.26}$. The actual  force is computed as an average over 
all realizations: $f_{\mathrm C}(\beta,L,h)=\frac{1}{N_{r}}\sum_{j=1}^{N_{r}}f_{\mathrm C}^{j}(\beta,L,h)$. 
We shall investigate  $f_{\mathrm C}(\beta,L,h)$ as a function of the two scaling variables $x=(L/\xi_0^+)^{1/\nu}t$
and $\hat w=h/L^{0.26}$ (see  Eq.~(\ref{eq:scal_force_0}):
\begin{equation}
 \label{eq:scal_force}
f_{\mathrm C}(\beta,L,h) \simeq L^{-3}\vartheta(x=(L/\xi_0^+)^{1/\nu}t, \hat w).
\end{equation}

For the pure system, i.e., $h=0$ we have also obtained  $N_{r}$ 
statistically independent values of the force $f_{\mathrm C}^{j}(\beta,L,h=0)$.
After averaging over the random surface fields for fixed values of  $L$ and of the inverse temperature $\beta$,
 we obtain the difference $\Delta f$
between  the force corresponding to the  random surface field  $h$
and the corresponding force for a pure system (with  $(o,o)$ BC): 
\begin{eqnarray}
\label{eq:MC-diff}
\Delta f(\beta,L,h) & = & \frac{1}{N_{r}}
\sum \limits_{j=1}^{N_{r}}\left[
f_{\mathrm C}^{j}(\beta,L,h)-f_{\mathrm C}^{j}(\beta,L,h=0) \right] \nonumber \\
&\simeq & L^{-3}\left[\vartheta(x,\hat w)-\vartheta(x,\hat w=0) \right].
\end{eqnarray}

The  statistical error is  inferred from
 10 series of MC steps for every realization.
The variance of different realizations of the disorder field 
is slightly smaller than the statistical error of a given realization.
The  error bars shown take into account only the statistical error.

\subsection{Numerical results}
\label{subsec:numres}

 First we  check whether, similar to the pure case  with nonzero surface fields $H_1$,
there is a nontrivial  dependence of the critical Casimir amplitude on the strength $h$ of the disorder.
As mentioned  above, for symmetric films
 the (negative) critical Casimir amplitude as a function of the   non-random surface field $H_1$ 
varies from its  value at  $H_1=0$  (ordinary transition fixed point)
to its   value at $H_1 = \infty$  (normal transition fixed point) 
in a non-monotonous way, i.e., 
through a maximum located at $L \simeq \ell_{ord}$ \cite{MCD,vas-11}. 
(In $d=3$, the absolute value of the critical
Casimir amplitude  for  $(o,o)$ BC is smaller than the one  for $(+,+)$ BC,  whereas in $d=2$  
they are equal.) 

\begin{figure}
\includegraphics[width=\textwidth]{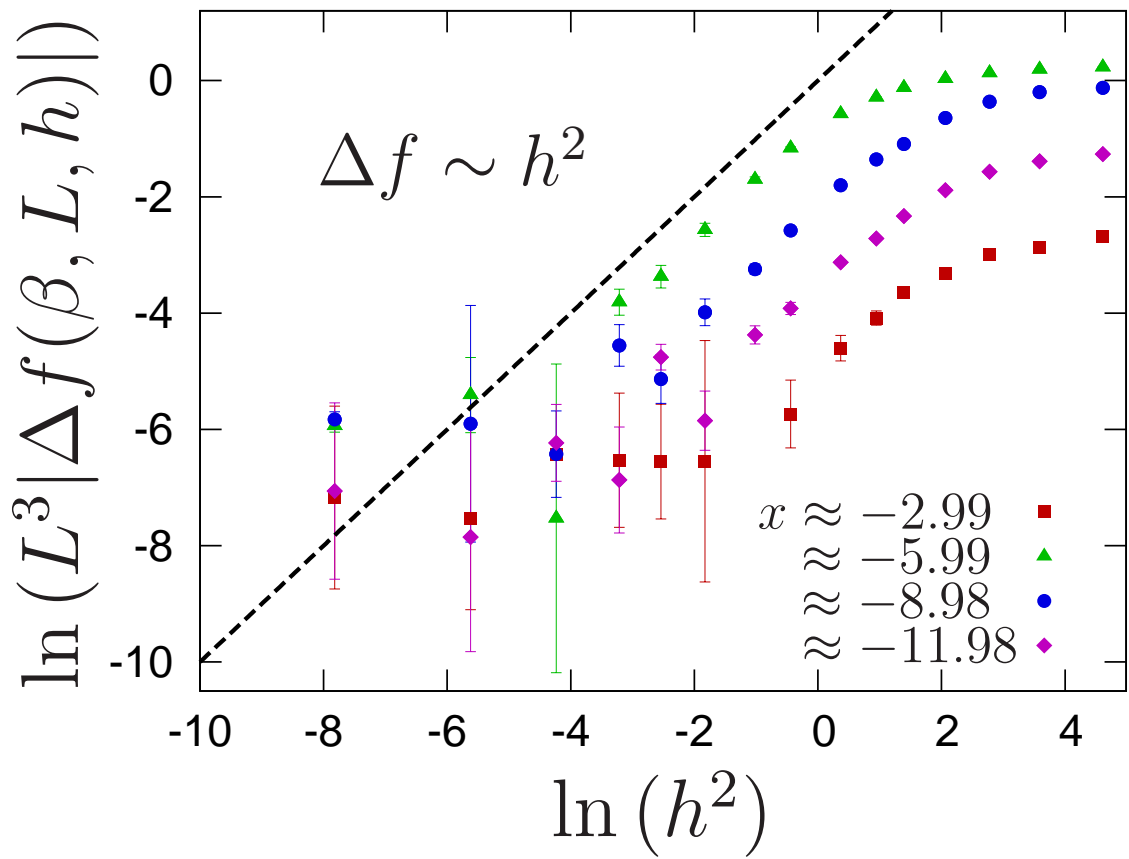}
\caption{ (Color)
 Log-log plot of the absolute value of the rescaled force difference   $L^{3}|\Delta f(\beta,L,h)|$ as  function 
of $h^{2}$  (Eq.~(\ref{eq:MC-diff}))
for several values of the temperature scaling variable  $x = (L/\xi_0^+)^{1/\nu}t \approx  -2.99,  -5.99, -8.98, -11.98$
and for  the system size $L_{z} = L  +\frac{1}{2} = 15$.
The straight dashed line indicates the slope corresponding to the proportionality  $\Delta f \sim h^2$.
The amplitude $\Delta f/h^2$ varies as function of $x$.
}
\label{fig:1}
\end{figure}

In the case of disorder, at $h=0$ the critical Casimir amplitude is  the one for  $(o,o)$ BC.
In analyzing our data for nonzero values of $h$ we have observed that  at the bulk critical point the difference $\Delta f$ is vanishingly small.
On the other hand, below $T_c$ (around  $x = (L/\xi_0^+)^{1/\nu}t \simeq -7$)  it exhibits  a  pronounced minimum.
Therefore, instead of considering the dependence  on $h$ of the critical Casimir amplitude 
we have studied  the critical Casimir force difference $\Delta f$ as a function of $h$
  for several fixed values of the temperature scaling variable around the minimum,  i.e., 
$x \approx  -2.99,  -5.99, -8.98, -11.98$. We have considered $h\in [0,5]$ and have found that, upon increasing $h$,
 $|\Delta f|$ increases {\it monotonically} with $h$  from  0  to a certain  $x$-dependent saturation value at large $h$.
Such a  leveling off is expected to occur, as discussed in Sec.~\ref{sec:rs}. 
In contrast, in the pure case, the small absolute  value of the Casimir 
amplitude $|\Delta_{(o,o)}|$ at first {\it decreases} even further upon increasing $H_1$ from zero,  reaches a minimum,
and only then increases towards the large value $|\Delta_{(+,+)}|$ for $(+,+)$ BC \cite{MCD,vas-11}.
For small values of $h$,  $\Delta f$ can be  described well by 
 a quadratic function of $h$. A crossover from the quadratic dependence  to 
 saturation of $\Delta f$ as a function of $h$
  occurs above  $\ln h^2 \approx -0.5$ (i.e., $h \approx 0.78$), corresponding to $\ell \approx 0.38$ so that  $L/\ell \approx 38$. 
  The leveling off occurs for $h \approx 5$, corresponding to $\ell \gg L$ 
 (see  Fig.~\ref{fig:1}).

As the next  step we determine the scaling function $\vartheta(x,\hat w)$.
Due to the finite and rather limited sizes of the lattices, which can be studied via MC simulations 
with presently available resources, one cannot expect to
reach the asymptotic regime where the true finite-size scaling holds. 
In order to obtain data collapse and thus being able to infer the leading universal scaling functions, one has to apply 
corrections  both to the scaling function and to  the scaling variables. These corrections to scaling 
are nonuniversal; they  depend on  details of the model as well as
on  the geometry and on the boundary conditions \cite{PF,Luck}.
Besides the bulk corrections to scaling, also  surface and finite-size ones can occur  \cite{privman}. 
Three-dimensional slabs of thickness $L$ exhibit a phase transition of two-dimensional character at a shifted critical
point $T_c(L)$ with $T_c(L\to \infty) - T_c(\infty) \sim L^{-1/\nu}$. For temperatures near $T_c(L)$
 one faces  considerable finite-size corrections due to the finite lateral system size $L_x = L_y = L_{||}$.
 This  leads  to a dependence of the critical Casimir forces on  the aspect ratio $\rho = L_z/L_{||}$.
 In the case of  periodic BC in the normal direction this dependence
 is  strong for  $\rho > 1/2$ \cite{HGS}.
 Here, we
 focus  on the film geometry (i.e., $\rho \to 0$), which can be realized in fluid systems by, e.g., wetting films.
 As it follows from our previous studies \cite{VGMD}, for small $\rho$  the scaling function of the critical Casimir 
 force with $(o,o)$ BC does depend on the aspect ratio but only within a certain  interval near its minimum.
 Here we take  $\rho = 1/6$ and  
neglect  the aspect ratio corrections.

On the other hand, we incorporate those corrections to scaling, which are due to the finite size $L$ in
normal direction.
In the present study, the following quantities are expected to acquire
corrections to scaling:
\begin{itemize}
\item  the amplitude of the scaling function $\vartheta =L^{3} f_{C}$
\item  the random surface field scaling variable $\hat w$
\item  the temperature scaling variable $x =\left(L/\xi_{0}^{+} \right)^{1/\nu}t$.
\end{itemize}

One may   expect that the scaling function  of the critical
Casimir   force $f_{C}$ additionally depends on $L^{-\omega'}$: 
$
 f_{C}(\beta,L,h) = L^{-3} \theta(x,\hat w, L^{-\omega'}) \simeq L^{-3} \vartheta(x,\hat w) [1 +L^{-\omega'} \zeta(x,\hat w)  + \ldots]
$
for $ L\gg 1$.
(Recall that actually $f_C$ depends also on the lattice spacing $a_0$ so that the correction to scaling scales as $(L/a_0)^{-\omega'}$; we set $a_0=1$.)
The  exponent $\omega'$  controls the leading
correction to the scaling behavior of the lattice estimate $f_{C}$. 
In the presence  of boundaries,  two main  corrections to scaling are expected to occur.
One is due to the  irrelevant bulk scaling fields \cite{Wegner}, which introduce exponents $\omega_i$
which cannot be expressed in terms of  the usual critical exponents such as  $y_t$ and $y_b$. 
The latest estimate for the value of the  smallest, and thus most relevant, of those exponents is $\omega \approx 0.832(6) $ for the $d = 3$ 
Ising model \cite{Hasenbusch82}.  
The other  correction terms stem from  the  irrelevant {\it surface}  contribution to the
Hamiltonian ${\mathcal  H}$ ~\cite{binder}. The value of $\omega'$ is determined by 
that irrelevant surface or bulk scaling field which has the smallest scaling dimension and which also
affects $f_{C}$. 
 The influence of the bulk corrections to scaling can be 
reduced by using improved Hamiltonians and
observables, which can also serve as representatives of the same universality
class. This is described in detail in Ref.~\cite{PV}. 
The value of the  exponent, corresponding to the least  irrelevant surface contributions in our system, is not known.
Even if the bulk correction-to-scaling exponent is dominant, by fitting the data it is difficult to disentangle
corrections $\propto L^{-\omega}$ and $\propto L^{-1}$. The latter can occur due to the presence of the boundaries. 
Moreover, for small lattice sizes,
next-to-leading corrections to scaling
might also be numerically important, resulting in effective  exponents.
The current accuracy of our MC  data and the range of sizes $L$ investigated here
do not allow for the
reliable determination of $\omega'$,  the function $\zeta(x,\hat w)$, and   the effective exponents. 
The analysis of various observables \cite{hucht}  revealed that also $x$ 
acquires a leading Wegner correction \cite{Wegner} of the form 
$
x \equiv \tau (L/\xi_0^+)^{1/\nu} (1 + g_\omega L^{-\omega}).
$
A detailed analysis of all types of corrections is beyond the scope of the
present study and is left to future research.

In our previous MC simulations aimed at obtaining 
 critical Casimir forces  for Ising  films
with a variety of {\it universal}  boundary conditions \cite{binder}, such as $(+,+), (+,-)$, 
or $(o,o)$ BC  ~\cite{VGMD,vas-11}, corrections to scaling were taken into account in an effective way
by using various ans\"atze. The choice  for a particular form
of corrections to scaling was guided by achieving 
the best  data collapse or the best fits.
With the lack of corresponding  theoretical  guidance, in the present study we have adopted the same, pragmatic approach.
First, as a phenomenological ansatz for the effective corrections we take 
$\omega' \simeq \omega \simeq 1$.
Second,  we follow a well established procedure of incorporating corrections to scaling  by introducing  
an  effective thickness $L+\hat\delta$ \cite{Hasenbusch,Hasenbusch84,vas-11,Hasenbusch15}. Accordingly, our ansatz for the corrections
to scaling is
\begin{equation}
\label{eq:ff}
 f_{C}(\beta,L,h)=
 (L+\hat\delta(h))^{-3}\vartheta(\left[
(L+\hat\delta(h))/\xi_{0}^{+}\right]^{1/\nu}t,\left(L+\hat\delta(h)\right)^{-1/(y_1 -(d-1)/2)}h).
\end{equation}
(We note that   this way the leading  bulk correction 
to scaling is treated  ``effectively'', because   the  correction of the  the scaling function  
  $(L+\hat\delta(h))^{-3}$  has the expansion
  $L^{-3}(1-3\hat\delta(h)/L+.... )$.)
The nonuniversal length $\hat\delta(h)$ is  fixed in such a way that the data scatter  due to
different $L$ is minimal.
In order to employ such a  correction-to-scaling scheme the knowledge of 
the whole surface  $\vartheta(x,\hat w)$ is needed, which is computationally
too demanding (see the Appendix in Ref.~[14(b)] where   we have discussed in detail our  strategy 
of obtaining the best fit for the values of the parameters which control the corrections to scaling).
 In our simulations we have generated data which correspond to only several cuts
 of the surface $\vartheta(x,\hat w)$
along $\hat w =const$.
Therefore we apply corrections to scaling along these cuts by  introducing 
$\delta(\hat w)$ and the effective thickness
$L_{\mathrm{eff}}(\hat w) = L+\delta(\hat w) = 
L_{z}-0.5+\delta(\hat w)$.
Accordingly, we introduce  a corrected scaling variable
\begin{equation}
\label{eq:x1}
 y=y(x,\hat w,L)= \left( \frac{L_{\mathrm{eff}}(\hat w)}{\xi_{0}^{+}}\right)^{1/\nu} t =
 \left( \frac{L_{\mathrm{eff}}(\hat w)}{L}\right)^{1/\nu} x = \left(1+\frac{\delta(\hat w)}{L}\right)^{1/\nu}x.
\end{equation}
Plotting  $L^3_{\mathrm{eff}} f_C(\beta,L,h)$ versus $y$ for fixed $\hat w$ and choosing $\delta(\hat w)$ 
such that data collapse is 
promoted for large $L$, one obtains a scaling function $g(y,\hat w)= L^3_{\mathrm{eff}} f_C(\beta,L,h)$, which for large $L$
does not exhibit  an explicit dependence on $L$. (This is achieved for smaller values of $L$  than for the scaling leading 
to the scaling function  $\vartheta(x,\hat w)$ introduced before by considering $L^3f_C(\beta,L,h)$.)
From the knowledge of the scaling function  $g(y,\hat w)$ one can construct the desired scaling function $\vartheta(x,\hat w)$
according to $\vartheta(x,\hat w) = g(y=x, \hat w)+ O(\delta(\hat w)/L)$.

The nonuniversal parameters $\delta(\hat w)$ are fixed in such a way that the data collapse of the 
Monte Carlo data for $L=10,15$, and 20 is optimal 
in the region $-10<  y <-2$.
Our corresponding results for $\delta(\hat w)$ are presented in Table~\ref{tab:fit1}.

\begin{table}[h]
\caption{Correction-to-scaling parameter 
 $\delta(\hat w)$ (see Eq.~(\ref{eq:x1}))}
 \begin{center}
\begin{tabular}{|c|c|c|c|c|c|c|c|}
\hline 
 $\hat w$& & 0 & 0.25 & 0.5 & 0.75 & 1 &  2 \\
\hline
 $\delta(\hat w)$ & & 1.6(1)  & 1.7(1)     & 1.67(10)    & 1.4(1)     & 1.05(10)   &  0.1(1)   \\
\hline
\end{tabular}
\end{center}
\label{tab:fit1}
\end{table}

The results of the above  procedure depend on the range of $y$  considered for the data collapse.

\begin{figure}
\includegraphics[scale=0.8]{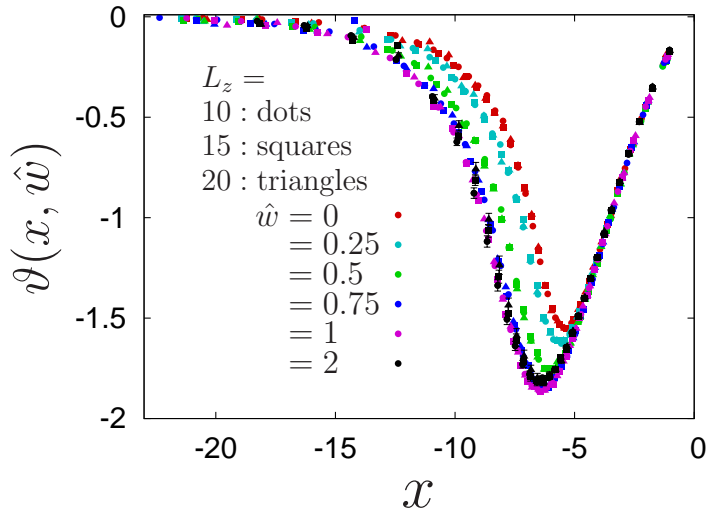}
\caption{ (Color)
Scaling function $\vartheta(x,\hat w)$ (Eq.~(\ref{eq:scal_force})) of the  critical Casimir force 
  for $3d$ Ising slabs  with random surface fields for several values of the random surface 
 field scaling variable  $\hat w = 0, 0.25, 0.5, 0.75, 1$ and 2 (from top to bottom).
 This scaling function has been obtained according to the procedure described in the main text.
 The MC data reported in this figure refer to 
 slabs with thicknesses  $L_{z}=10$ (dots), $15$ (squares), $20$ (triangles) and indicate that for fixed
 $\hat w$ data collapse has been accomplished.  
For $ \hat w\le 0.75$ the error bars are smaller than the symbol size. 
The pure case $\hat w=0$ is also shown (see Ref.~\cite{VGMD}).
}
\label{fig:2}
\end{figure}

By applying the  rescaling procedure as described above, we have obtained an estimate for 
the scaling function $\vartheta(x,\hat w)$ of  the   Casimir force  $f_C(\beta,L,h)$ for 
the $3d$ Ising model in  the slab geometry with random surface fields.
It is shown in 
Fig.~\ref{fig:2} as a function
of the  scaling variable $x$
 for the values 
 $\hat w=0,0.25,0.5,0.75,1,$ and 2.
One sees  that the rescaling procedure leads to  data collapse  for  various system sizes $L$,
for  a {\it given} value of $\hat w$. 
However,  the rescaled data for {\it  different} values 
of $\hat w =0, 0.25, 0.5$ do not collapse. The curves for $\hat w = 0.75, 1$ and 2 lie almost on the top of each other.
Note that  $\hat w = 0.75$ and 1  lie in the crossover  regime to the ``strong disorder limit'', where 
 $L <  \ell$, whereas for  $\hat w = 2 $   this limit is almost attained
(compare Fig. 1).
For strong disorder, as discussed earlier,  the  majority of  the surface spins  do not fluctuate but are frozen 
to the values -1 or  +1 
so that  as a function 
of $h^2$  the  contribution $\Delta f$  to the Casimir scaling function due to the random surface fields
 levels off, i.e., in this limit the value of $h$  does not matter.

\begin{figure}
\includegraphics[scale=0.8]{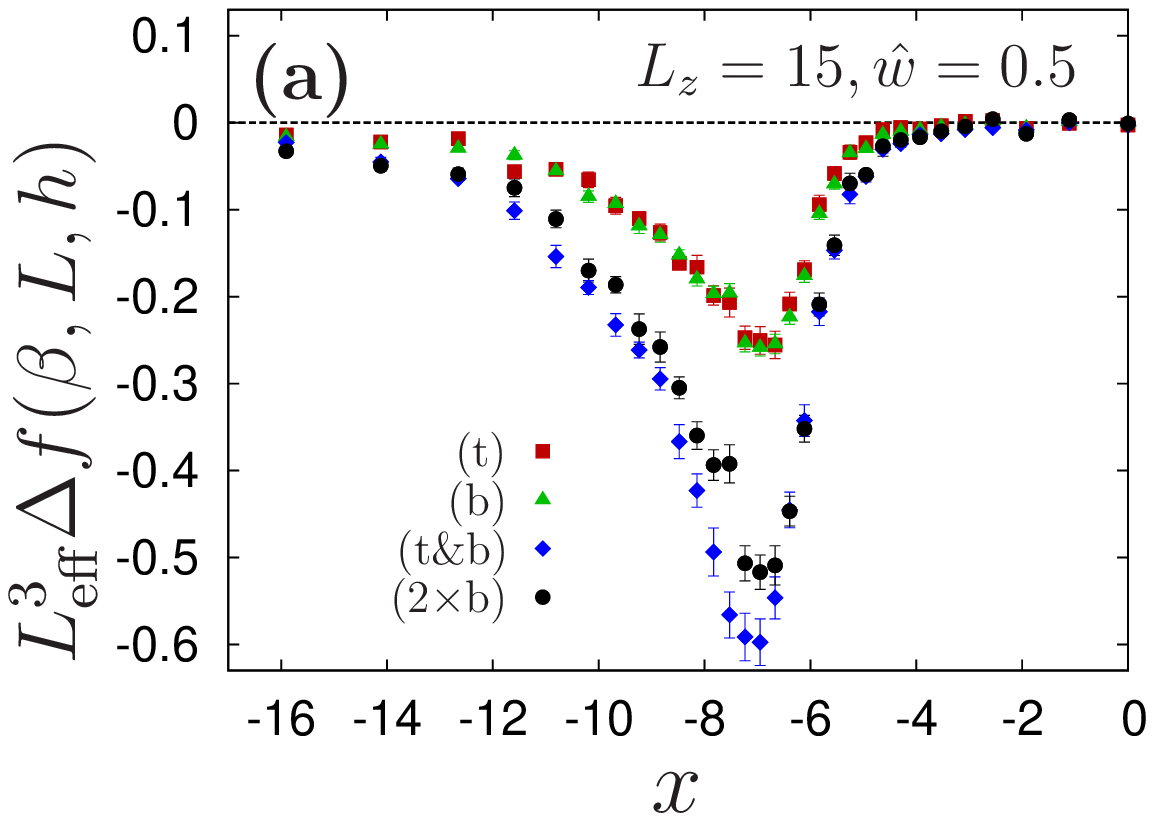}
\includegraphics[scale=0.8]{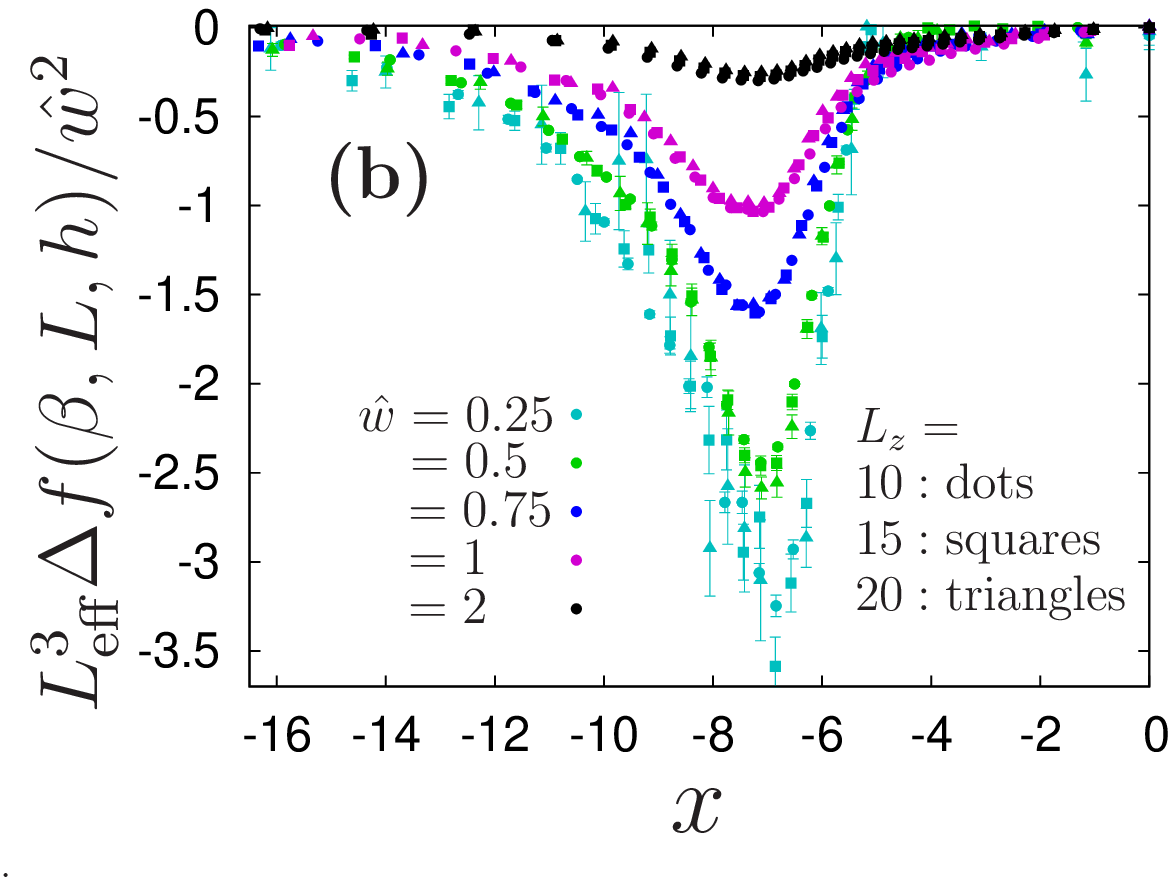}
\caption{(Color)
 (a) 
Additivity check: rescaled difference
$L^3_{\mathrm{eff}}\Delta f(\beta,L,h)$ as  function of the scaling variable 
$x$ (see Eq.~(\ref{eq:MC-diff}) and  main text)
for  the system size $L_{z}=15$ and $h=1.01(8)$ corresponsing to 
 $\hat w=0.5$. Random surface fields present only   at the top (t), only at the  bottom (b) (which has to yield identical results),
and on both sides of the system (t \& b). The data for bottom random fields 
multiplied by 2 are shown  for  comparison (note that $2 \times$b\,$\equiv$\,(t)+(b)). 
The difference between the black and blue data highlights the lack of additivity which leads to stronger forces.
In all plots $\hat w= h/L^{0.26}$.
(b) Rescaled difference, between the case of the presence and of the absence of randomness,
 of the  Casimir force scaling function
  $L^3_{\mathrm{eff}}\Delta f(\beta,L,h)/\hat w^2$  of the $3d$ Ising model in the slab geometry 
as a function of the scaling variable $x$.
The data correspond to the same thicknesses $L_z$  as in Fig.~\ref{fig:2}. The curves correspond
to $\hat w = 0.25, 0.5, 0.75, 1, 2$ (from bottom to  top). For $\hat w>0.75$ the error bars are smaller than the symbol size.
}
\label{fig:3}
\end{figure}

In order to gain more insight into the effect of  random surface fields on our system,
we use the estimate for $\vartheta(x,\hat w) $ shown in Fig.~\ref{fig:2} and calculate for each value of $\hat w$
the difference $\Delta f$ according to Eq.~(\ref{eq:MC-diff}). This is done 
by subtracting  from the
curve corresponding to the particular  value of
$\hat w$ the one corresponding to the pure case of $\hat w =0$ (red curve).
In Fig.~\ref{fig:3}(a) we show the result of this operation rescaled by $L^3_{\mathrm{eff}}$
 for an  Ising slab with  $L_{z}=15$ and for a random surface field  $h =1.01(8)$ corresponding to the 
 scaling variable $\hat w=0.5$ for three cases:
(i)  random surface fields applied only on the top side of the film (t),
(ii) only on the bottom side (b), and  (iii) on both sides (t \& b).
Obviously the results for the (t) and (b) cases coincide.
For comparison we have plotted  also the results for the bottom side multiplied by two ($2\times$b $\equiv$ (t)+(b)).
If the effects from the top and the  bottom sides were additive,
(t \& b) should coincide with ($2\times $b), which is not the case; the actual force is stronger. 
Thus we conclude that  the contribution to the CCF stemming from the random fields
 at both confining surfaces is {\it not} the sum of the contributions due to the random surface fields being present only
 at one of the two surfaces.

Finally, in order to  focus on the leading behavior of the  difference $\Delta f$ 
between the Casimir scaling 
functions for a system with and without disorder,
in Fig.~\ref{fig:3}(b)  we plot
$
L^3_{\mathrm{eff}}\Delta f(\beta,L,h)/\hat w^2
$
 for various values of $L_z$
and $\hat w$ as  function of the scaling variable $x$. 
For $\hat w = 0.25$ and 0.5, we observe  good data collapse within the error bars  of our data, confirming that as function of $\hat w$ the leading
behavior of the  difference $\Delta f$  for small $\hat w$ is quadratic in $\hat w$, 
consistent with the results shown in Fig.~\ref{fig:1}.
Based on these observations and  strengthened  by the symmetry property $p(H_1) = p(-H_1)$
of the surface field distribution, 
we put forward the hypothesis that for small $\hat w$  one has
\begin{equation}
\label{eq:scal_force_w}
 f_{\mathrm C}(\beta,L,h) \simeq L^{-3}\left\{\vartheta((L/\xi_0^+)^{1/\nu}t, \hat w=0) + \hat w^2 \vartheta_0((L/\xi_0^+)^{1/\nu}t)\right\}
\end{equation}
where $\vartheta((L/\xi_0^+)^{1/\nu}t, \hat w=0)$ is the scaling function of the critical Casimir force for $(o,o)$ BC. 
 The universal scaling
function $\vartheta_0(x)$  depends on  $x$ only; $\vartheta_0(x)$  is given by the curve in Fig.~\ref{fig:3}(b) formed by the  data corresponding to
 $\hat w =0.25$ and 0.5.
We note that Eq.~(\ref{eq:scal_force_w}) agrees with the expansion in Eq.~(\ref{eq:f_C_exp}) 
with a vanishing leading  correction-to-scaling
term.
From this we infer that the randomness induced  occurrence  of the extra contribution $\Delta f$ 
to the critical Casimir force  is due to the irrelevant scaling 
field $h$ (or more generally ${\tilde g_1} = h/c^{y_c}$). 
Moreover, at large $h$ such that  $\ell \ge L$ we observe  surface spin configurations of randomly distributed frozen spins 
 (see the discussion ath the end of Sec.~\ref{sec:rs}). For the corresponding CCF one has
\begin{equation}
\label{eq:scal_force_wh}
 f_{\mathrm C}(\beta,L,h) \simeq L^{-3}\vartheta((L/\xi_0^+)^{1/\nu}t, \hat w=\infty) =  L^{-3} \vartheta_w((L/\xi_0^+)^{1/\nu}t),
\end{equation}
where the scaling function $\vartheta_w(x)$ is approximately given by the curve in  Fig.~\ref{fig:2}  which is common to the data 
points corresponding to
$\hat w = 1$ and 2.
We note that for large $h$ the scaling analysis, which leads to the conclusion that the disorder
is an irrelevant perturbation of the ordinary surface universality class, does not hold.
Our findings that CCF in films are significantly influenced by the surface disorder for $L/\ell \approx 1$ should 
actually be  valid not only for thin slabs but for all slabs 
  in the scaling  limit $L \gg 1$ and $\ell =  \left({\tilde g_1}\right)^{-1/(y_1 -(d-1)/2)} \gg 1$
  such that $L/\ell$ is kept nonzero and finite, 
 which requires large values of ${\tilde g_1}$.
The present limits of the  accuracy of our data  do not allow us to draw quantitatively reliable 
conclusions concerning the behavior  of $\Delta f$ above $T_c$.

\section{Conclusions}
\label{sec:concl}

The  MC simulation  data  show that the presence
of random surfaces fields  with zero mean increases substantially the strength of  the critical Casimir force  as compared with  
the pure case without fields. The strongest effects occur when the length scale  $\ell $
associated with the random surface field
becomes comparable with the thickness of the film.
For  weak disorder this effect is proportional to
the square of  the strength of the random surface fields.  For strong disorder, the dependence  of the CCF on $h$ 
levels off.
For all  strengths  of  disorder,  at bulk criticality the  CCF decays asymptotically 
 as function of the film thickness $L$ 
as $L^{-3}$, which is the same behavior as for  the pure system.

 As mentioned in the Introduction, in the study of Ref.~\cite{francesco}
 the quenched random disorder is applied only to one of the two surfaces.
 Moreover,  different from the present work, it is governed by the 
   binomial distribution, i.e., spins on the surface  subjected to disorder 
 take  the value 1 with the probability $p$ and -1 with the probability $1-p$. On the other surface the spins are
 fixed at the value +1 or -1.
For $p=0.5$, for which the  mean value
 of the surface fields  vanishes, the Monte Carlo simulation data for  CCFs presented in Ref.~\cite{francesco} scale 
with the  inverse third power of the (effective)  film thickness. This is interesting, because 
the binomial distribution used in Ref.~\cite{francesco} represents  a sort of "strong disorder" limit; nevertheless,
the critical behavior is still governed by the ordinary fixed point. This is consistent with our findings.
For other values of $p$ considered in Ref.~\cite{francesco} a substantial dependence of the scaling function on $L$ is observed.
As  suggested by the author,   this latter  lack of  data collapse may be due to the negligence of
the scaling variable associated with the random surface field
 describing  the crossover from the ordinary to the normal phase transition.
 In our case, taking into account  the scaling variable connected with the random surface field $h$ was necessary in order to 
achieve data collapse.

 Our theoretical predictions lend themselves to be investigated  experimentally and pose a challenge to further analytic studies.
 The model studied here can be realized experimentally by confining a near-critical binary mixture such as lutidine-water 
 by two planar walls, each of them patterned by chemical stripes with alternating, strong preferences for the two species of the binary liquid
 mixture. In the limit of narrow stripes these surfaces mimic the ordinary surface universality class \cite{TZGVHBD}. The random surface
 fields can be realized by randomly adsorbing on these surface structures a binary mixture of adsorbates with opposite preferences
 for the two liquid species. The critical Casimir force can be obtained by AFM where the two surfaces are those of two crossed cylinders
 with large radii of curvature \cite{israel}.

\end{document}